\documentclass[manuscript]{emulateapj}

\shorttitle{Confined Solar Flares Observed by ONSET and SDO}
\shortauthors{Yang \& Zhang}

\journalinfo{Accepted for publication in ApJL}
\submitted{ApJL, accepted 2018 June 5}

\begin{document}

\title{Mini-filament Eruptions Triggering Confined Solar Flares \\ Observed by ONSET and SDO}

\author{Shuhong Yang\altaffilmark{1,2}, and Jun Zhang\altaffilmark{1,2}}

\altaffiltext{1}{CAS Key Laboratory of Solar Activity, National
Astronomical Observatories, Chinese Academy of Sciences, Beijing
100101, China; shuhongyang@nao.cas.cn}

\altaffiltext{2}{School of Astronomy and Space Science, University
of Chinese Academy of Sciences, Beijing 100049, China}

\begin{abstract}

Using the observations from the Optical and Near-infrared Solar Eruption Tracer and the Solar Dynamics Observatory, we study an M5.7 flare in AR 11476 on 2012 May 10 and a micro-flare in the quiet Sun on 2017 March 23. Before the onset of each flare, there is a reverse S-shaped filament above the polarity inversion line. Then the filaments become unstable and begin to rise. The rising filaments gain the upper hand over the tension force of the dome-like overlying loops and thus successfully erupt outward. The footpoints of the reconnecting overlying loops successively brighten and are observed as two flare ribbons, while the newly formed low-lying loops appear as the post-flare loops. These eruptions are similar to the classical model of successful filament eruptions associated with coronal mass ejections. However, the erupting filaments in this study move along large-scale lines and eventually reach the remote solar surface, i.e., no filament material is ejected into the interplanetary space. Thus both the flares are confined ones. These results reveal that some successful filament eruptions can trigger confined flares. Our observations also imply that this kind of filament eruptions may be ubiquitous on the Sun, from active regions with large flares to the quiet Sun with micro-flares.

\end{abstract}

\keywords{Sun: activity --- Sun: filaments, prominences --- Sun: flares --- Sun: magnetic fields}

\section{Introduction}

Solar flares, suddenly releasing a large amount of magnetic energy, are one of the most energetic phenomena on the Sun (Priest \& Forbes 2002; Janvier et al. 2015; Yang et al. 2017). Solar filaments are deemed to play a crucial role in the initiation of solar flares. For the onset of filament eruptions, several different mechanisms have been proposed and studied, such as kink instability, torus instability, flux emergence and cancellation, tether cutting reconnection, breakout reconnection (e.g., Antiochos et al. 1999; Chen \& Shibata 2000; Zhang et al. 2001; T{\"o}r{\"o}k \& Kliem 2005; Chen et al. 2016; Wyper et al. 2017). In the popular flare model and relevant observations, a rising filament stretches the overlying magnetic field lines, creating a current sheet between the anti-parallel field lines where magnetic reconnection takes place, and then a bulk of plasma and magnetic structure is ejected into the interplanetary space, forming a coronal mass ejection (CME; Svestka \& Cliver 1992; Shibata et al. 1995; Lin \& Forbes 2000; Ding et al. 2003).

However, some flares termed confined flares are not accompanied by CMEs (e.g., Guo et al. 2010; Zheng et al. 2012; Yang et al. 2014; Yan et al. 2015; Joshi et al. 2015). The main mechanism to explain the formation of confined flares is the failure of filament eruptions (Ji et al. 2003). Solar eruptions are greatly affected by the surrounding coronal structures. Simulations and calculations reveal that the strong overlying arcades can prevent magnetic energy release and result in confined flares (T{\"o}r{\"o}k \& Kliem 2005; Fan \& Gibson 2007; Wang \& Zhang 2007). If the decrease factor of the overlying field is not high enough, the eruption of the filaments will fail, thus forming confined flares. Moreover, Myers et al. (2015) found in a laboratory experiment that, if the tension force of the guide magnetic field is strong enough, a torus-unstable flux rope also fails to erupt. Recently, Amari et al. (2018) studied the evolution of a flux rope with the overlying loops forming a confining cage. Their results revealed that the flux rope had insufficient energy to break through the confining cage even if its twist was large enough to trigger a kink instability, resulting in a confined flare. For a failed filament eruption, the filament first accelerates, then decelerates, and eventually reaches the peak height, after which the filament drains back to its initial location (Ji et al. 2003; Cheng et al. 2015).

In this Letter, based on the observations from the Optical and Near-infrared Solar Eruption Tracer (ONSET; Fang et al. 2013) and the Solar Dynamics Observatory (SDO; Pesnell et al. 2012), we present the analysis of two confined flares associated with jet-like eruptive activity in the closed-field corona, referred to as confined jets in the simulations by Wyper \& DeVore (2016) and Wyper et al. (2016). ONSET, a new multi-wavelength solar telescope of Nanjing University, is installed at Fuxian Solar Observatory in Yunnan Province of China. It can simultaneously image the photosphere, chromosphere, and corona in white-light (3600 {\AA} and 4250 {\AA}), H$\alpha$ 6562.8 {\AA}, and He I 10830 {\AA}.

\section{Observations and data analysis}

\begin{figure*}
\centering
\includegraphics
[bb=46 230 539 605,width=0.75\textwidth]{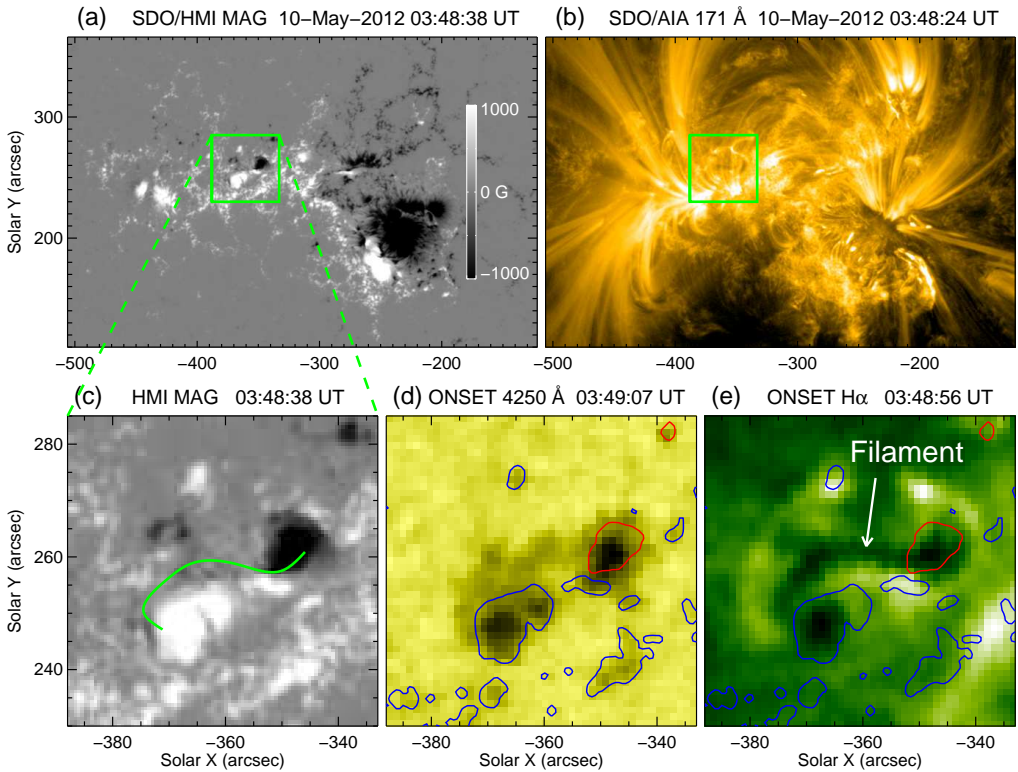} \caption{Panels (a) and (b): HMI LOS magnetogram and AIA 171 {\AA} image showing the overview of AR 11476. Panels (c)-(e): expanded HMI LOS magnetogram, ONSET 4250 {\AA} image, and ONSET H$\alpha$ image of the area where the M5.7 flare occurred. The green curve in panel (c) represents the filament identified in the H$\alpha$ image. The blue and red curves in panels (d) and (e) are the contours of the positive and negative magnetic fields at $\pm$ 600 G, respectively. \label{fig1}}
\end{figure*}

\begin{figure*}
\centering
\includegraphics
[bb=51 167 548 664,width=0.75\textwidth]{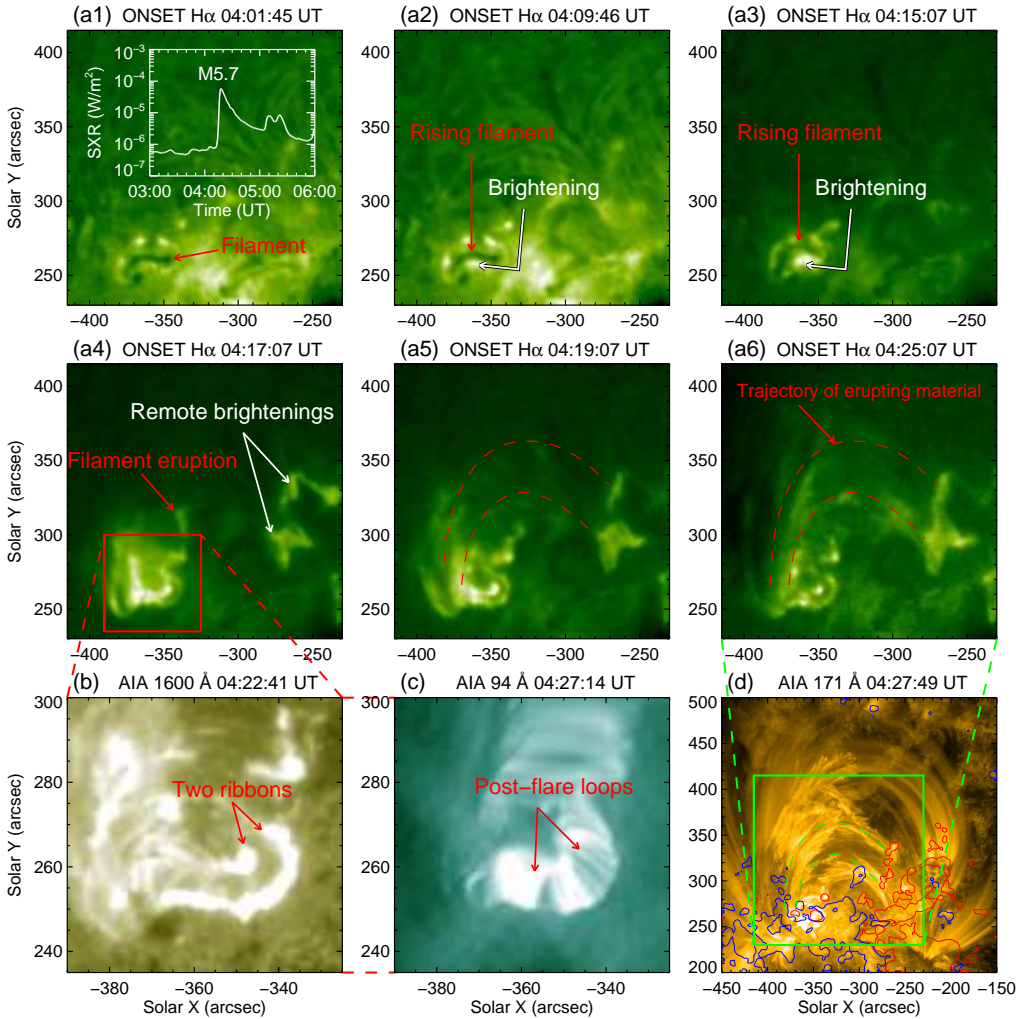} \caption{Panels (a1)-(a6): ONSET H$\alpha$ images displaying the process of the filament eruption. Panels (b) and (c): AIA 1600 {\AA} and 94 {\AA} images showing the two ribbons and post-flare loops due to the filament eruption. Panel (d): AIA 171 {\AA} image showing the large scale coronal structure above the flaring site. The white curve in panel (a) displays the variation of the GOES soft X-ray (1-8 {\AA}) flux, and the blue and red curves in panel (d) are the contours of the positive and negative magnetic fields at $\pm$ 100 G, respectively. \protect\\Two animations (Movie1.mp4 \& Movie2.mp4) of this figure are available. \label{fig2}}
\end{figure*}

\begin{figure*}
\centering
\includegraphics
[bb=46 251 539 605,width=0.85\textwidth]{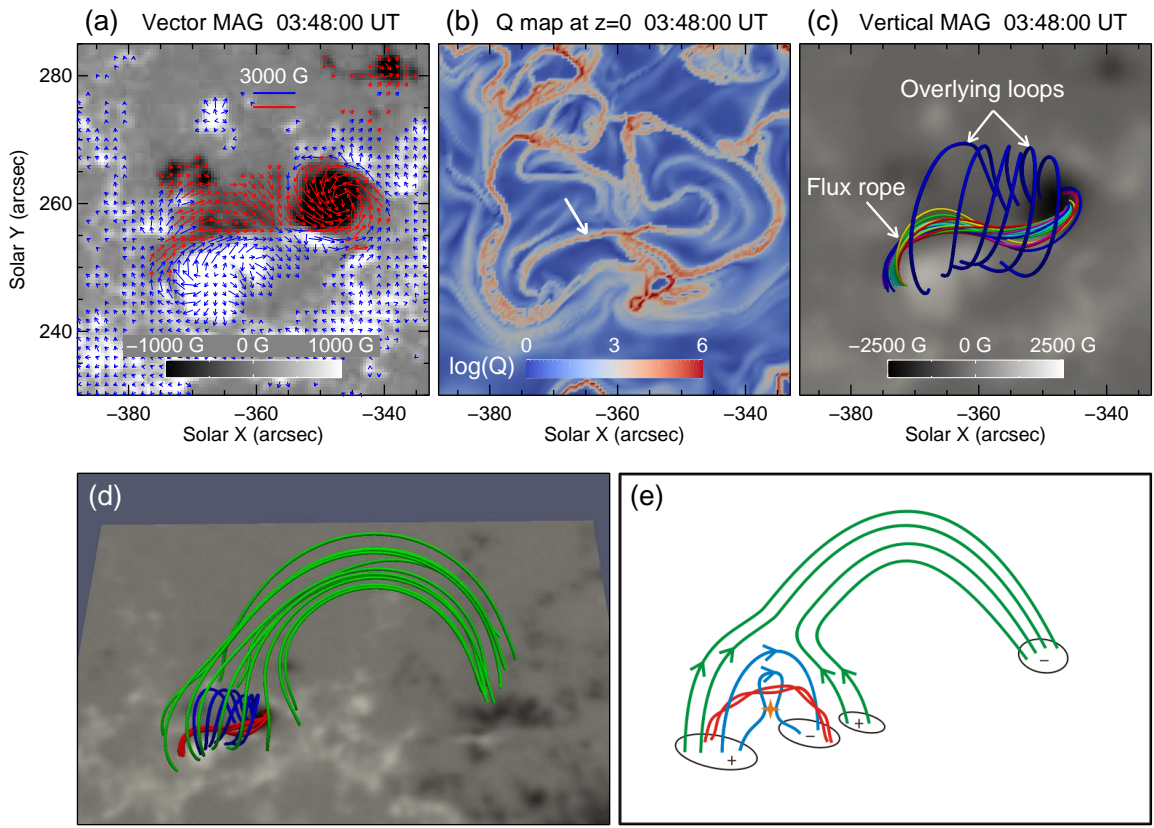} \caption{Panels (a) and (b): HMI vector magnetogram of the flaring region, and the corresponding squashing factor $Q$ map, respectively. Panel (c): top-view of the NLFFF extrapolated magnetic flux rope (with $\mathcal{T}_w > 1$) and overlying loops with the background of photospheric vertical field. Panel (d): side-view of the reconstructed coronal structures. Panel (e): sketch showing the mechanism of the confined flare resulting from the successful filament eruption. The red, blue, and green curves in panels (d) and (e) represent the magnetic flux rope, overlying loops, and outer large-scale field lines, respectively. The star symbol in panel (e) marks the reconnection site. \label{fig3}}
\end{figure*}

\begin{figure*}
\centering
\includegraphics
[bb=42 287 539 546,width=\textwidth]{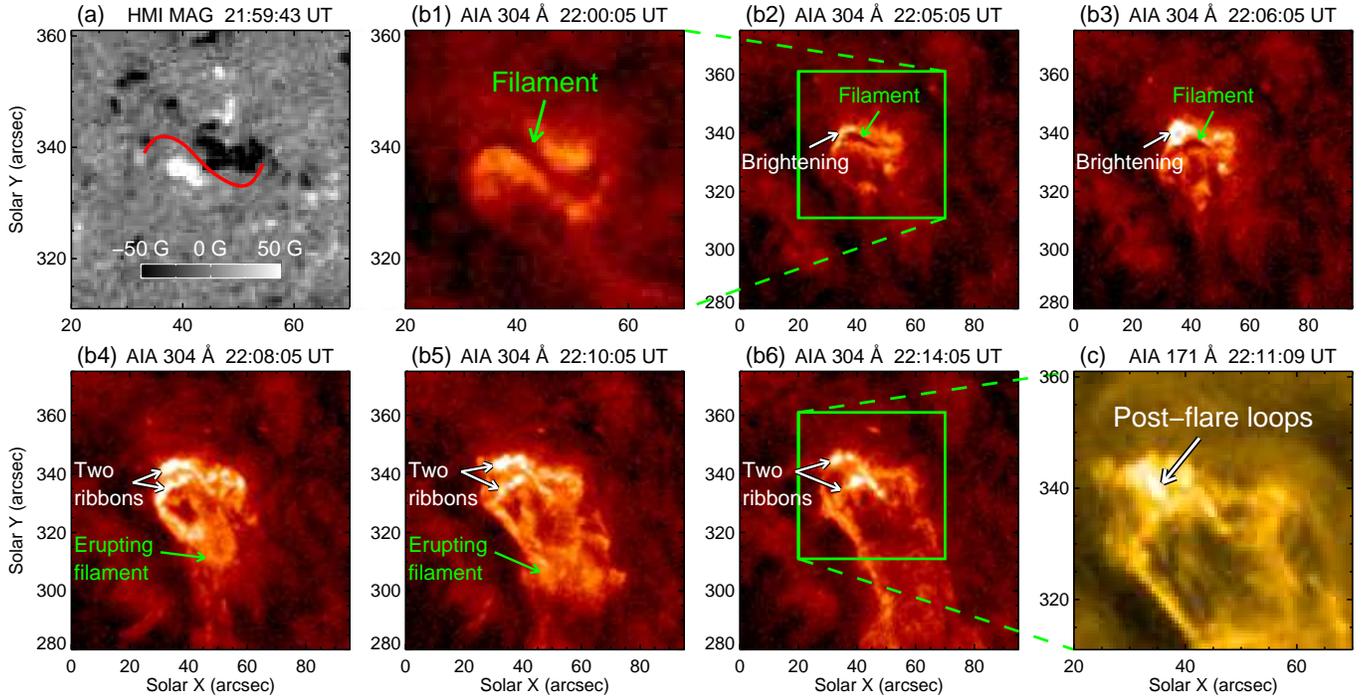} \caption{Panel (a): HMI LOS magnetogram on 2017 March 23 showing the photospheric magnetic field where the mini filament was located. Panels (b1)-(b6): AIA 304 {\AA} images displaying the eruption process of the filament. Panel (c): AIA 171 {\AA} image showing the post-flare loops due to the filament eruption. \protect\\One animation (Movie3.mp4) of this figure is available.\label{fig4}}
\end{figure*}

\begin{figure*}
\centering
\includegraphics
[bb=81 314 510 525,width=0.8\textwidth]{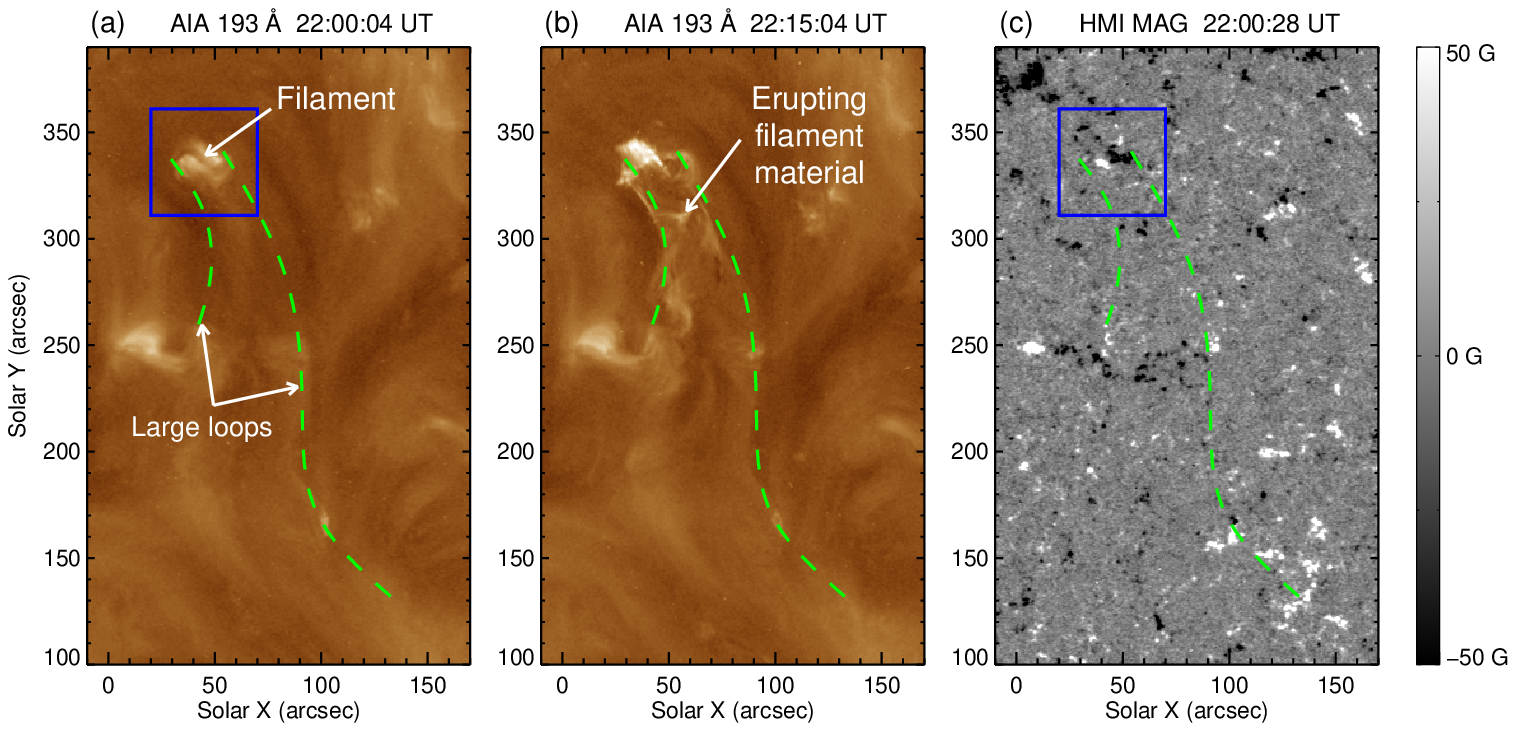} \caption{Panels (a)-(b): AIA 193 {\AA} images showing the large-scale coronal structures before and after the filament eruption, respectively. Panel (c): corresponding HMI LOS magnetogram displaying the underlying photospheric magnetic field. The green curves indicate the large-scale coronal loops along which the filament material reached the remote solar surface. The blue windows outline the FOV of Figures 4(a)-(b1). \protect\\One animation (Movie4.mp4) of this figure is available. \label{fig5}}
\end{figure*}

We mainly study an M5.7 flare in AR 11476 on 2012 May 10 and a micro-flare in the quiet Sun on 2017 March 23. The M5.7 flare on 2012 May 10 was observed by ONSET and SDO. We use the ONSET white light 4250 {\AA} and H$\alpha$ 6562.8 {\AA} full-disk images between 03:45 UT and 04:27 UT. They have a cadence of 1 min and a pixel size of about 1{\arcsec}. The simultaneous SDO/Atmospheric Imaging Assembly (AIA; Lemen et al. 2012) multi-channel images and SDO/Helioseismic and Magnetic Imager (HMI; Scherrer et al. 2012) line-of-sight (LOS) magnetograms are also adopted. The AIA UV/EUV full-disk images have a cadence of 24/12 s and a spatial sampling of 0.{\arcsec}6 pixel$^{-1}$, and the HMI LOS magnetograms have a cadence of 45 s and a pixel size of 0.{\arcsec}5. The Geostationary Operational Environmental Satellite (GOES) data with a 1 min cadence are used to examine the variation of soft X-ray flux. For the micro-flare on 2017 March 23, we mainly employ the AIA 304 {\AA}, 171 {\AA}, and 193 {\AA} images and HMI LOS magnetograms from 22:00 UT to 22:50 UT. All the AIA and HMI data are first calibrated using the standard routine within the Solar Software package, and then de-rotated differentially. The reference times of derotation for the M5.7 flare and micro-flare are 04:00 UT and 22:00 UT, respectively. The ONSET images are coaligned with the SDO data by cross-correlating specific features.

In order to study the magnetic topology of AR 11476, we use the HMI data product called Space weather HMI Active Region Patches (SHARP) of the photospheric vector magnetic fields. The photospheric vector magnetogram is preprocessed with the method of Wiegelmann et al. (2006) to best suit the force-free condition. Then the preprocessed magnetogram is used as the bottom boundary to reconstruct the coronal structures using the nonlinear force-free field (NLFFF) modeling (Wheatland et al. 2000; Wiegelmann 2004). The NLFFF extrapolation is performed in a box of 896$\times$512$\times$256 uniformly distributed grid points with $\Delta x= \Delta y=\Delta z=0.$\arcsec$5$. Moreover, we calculate the twist number $\mathcal{T}_w$ (Berger \& Prior 2006) and squashing factor $Q$ (D{\'e}moulin et al. 1996; Titov et al. 2002) of the extrapolated magnetic field using the code developed by Liu et al. (2016).

\section{Results}

The leading and trailing polarities of AR 11476 shown in Figure 1(a) are negative and positive, respectively. The region of interest is outlined by the green window, which is located nearby one leg of a cluster of large-scale coronal loops seen in 171 {\AA} (panel (b)). The zoom-in views of this region are shown in panels (c)-(e). We can see that, in the center of the FOV, there were mainly two magnetic concentrations with opposite polarities, forming a dipolar region (panel (c)). The positive and negative patches corresponded to a pair of sunspots in the photosphere, as seen in ONSET 4250 {\AA} image (panel (d)). In the ONSET H$\alpha$ image at 03:48:56 UT, there was a filament lying along the polarity inversion line (PIL). This area is just the site where the M5.7 flare took place.

The M5.7 flare started at 04:11 UT, peaked at 04:18 UT, and ended at 04:23 UT, with the duration of 12 min, as revealed by the variation of the GOES soft X-ray flux (the white curve in Figure 2(a1). Before the initiation of the flare (e.g., at 04:01:45 UT), the state of the filament was generally stable (see the H$\alpha$ image in panel (a1)). At 04:09:46 UT, the filament became unstable and began to rise, meanwhile a slight brightening appeared beneath the filament (panel (a2). The filament went on rising and the flaring brightness increased (panel (a3), also see the animation Movie1.mp4). Then the filament erupted outward and the remote area denoted by the white arrows brightened (panel (a4)). The erupting material moved along the trajectory outlined by the dashed curves in panels (a5)-(a6) and reached the remote brightening region. Due to the eruption of the filament, two flare ribbons can be clearly observed in the chromosphere, as shown in AIA 1600 {\AA} image (panel (b)), and a set of post-flare loops also appeared, as shown in AIA 94 {\AA} image (panel (c)). In the 171 {\AA} image with a large FOV (panel (d)), the erupting material flew along the large-scale loops, from the left leg to the remote one (see the animation Movie2.mp4).

In order to study the magnetic topology of the filament, we analyze the vector magnetic fields at the filament location. The HMI vector magnetogram at 03:48:00 UT is shown in Figure 3(a). The vortex patterns of the horizontal fields (blue and red arrows) at both the positive and negative polarities indicate that the magnetic fields are nonpotential and the field lines are greatly twisted. In the $Q$ map at the photosphere ($x-y$ plane with $z=0$), the main PIL of the magnetic fields coincides with a high $Q$ region (denoted by the white arrow in panel (b)), indicating the existence of a quasi-separatrix layer therein. In the NLFFF extrapolated fields, we find there is a flux rope with $\mathcal{T}_w > 1$ (indicated by the colorful wrapped lines in panel (c)) located above the PIL. This flux rope exactly corresponds to the filament identified in the H$\alpha$ observations. There also exist many overlying loops forming a dome-like structure (see the blue lines) above the flux rope. In the side-view of the reconstructed large-scale coronal structures (see panel (d)), the flux rope (red curves) and overlying loops (blue ones) are surrounded by large-scale field lines extending upward from the positive fields and ending in the remote negative fields (green curves). Based on the observations, we draw a cartoon to illustrate the successful eruption of the filament and the subsequent formation of confined flare, which will be addressed in Section 4.

In the quiet Sun, the successful eruption of a mini-filament as shown in Figure 4 led to a confined micro-flare (also see the animation Movie3.mp4), similar to the process of the M5.7 flare. In the small FOV of the quiet region (see panel (a)), the HMI LOS magnetogram at 21:59:43 UT presented a dipolar magnetic patches with the opposite polarities saturated at $\pm$ 50 G. Along the PIL, a reverse S-shaped mini-filament can be identified in the 304 {\AA} image (panel (b1)). At 22:05:05 UT, the filament began to rise and there was a brightening beneath the left part of the filament (see panel (b2)). Only 1 min later, the filament lifted significantly and the underneath brightening was more conspicuous, triggering a micro-flare. At 22:08:05 UT, the filament erupted outward (indicated by the green arrow in panel (b4)), and two obvious flare ribbons (denoted by the white arrows) appeared at the initial location of the filament. Afterward, the material of the filament went on flowing further, and the separation of the flare ribbons became larger (see panels (b5)-(b6)). In addition, many post-flare loops (indicated by the arrow in panel (c)) connecting the two ribbons can be well observed in AIA 171 {\AA} image.

At the pre-flare stage, the large-scale coronal structures observed in AIA 193 {\AA} are shown in Figure 5(a). From the underlying photospheric magnetogram (panel (c)), we can see that the surrounding fields of the mini-filament had a negative polarity. In AIA 193 {\AA} image, there were large-scale loops anchoring nearby the mini-filament and connecting to the remote positive fields (outlined by dashed curves). As the filament successfully erupted, the erupting material moved outward along the large-scale loops, as shown in panel (b). It is especially clearly that some filament material flew along the left loops (indicated by the left green curve in panel (b)) and reached the end with positive fields (also see the animation Movie4.mp4).

\section{Conclusions and discussion}

With the multi-instrument observations from ONSET and SDO, we have studied the evolution of an M5.7 flare in AR 11476 and a micro-flare in the quiet Sun. For each flare, there existed a filament above the PIL at the pre-flare stage. Then the filaments became unstable and began to rise with brightenings beneath them. The rising filaments gained the upper hand over the tension force of the dome-like overlying field lines and thus erupted successfully. Consequently, two flare ribbons and many post-flare loops were formed due to the filament eruption. The erupting filament material flew outward along large-scale lines with the other leg anchoring in the remote area. The filament material finally reached the remote end, and no material was ejected into the interplanetary space. These observations are consistent -- both in magnetic topology and overall eruption morphology -- with the simulation results from Wyper \& DeVore (2016) and Wyper et al. (2016), who referred to this kind of events as confined jets. Figure 17 in the paper of Wyper et al. (2016) showing a synthetic emission proxy displays exactly the kind of dynamics reported here, both from a side-view (comparable to the M5.7 event) and a top-down view (comparable to the quiet-Sun event).

For the M5.7 flare, it was also detected in the white light (Song et al. 2018). This flare was trigged by the filament eruption, and we use the sketch in Figure 3(e) to describe this process. The flux rope (i.e., the filament in observation) lost its stability due to some mechanisms, such as the increasing twist caused by the photospheric shear motion or the reconnection between the overlying loops (represented by the blue curves) and the anti-directed large-scale field lines (indicated by the right-side green curves). Then the rising flux rope stretched the overlying loops, and magnetic reconnection occurred at the site marked by the star symbol. Magnetic reconnection can convert magnetic energy into thermal energy and kinetic energy of solar plasma (Cirtain et al. 2013; Tian et al. 2014; Li et al. 2015; Yang et al. 2015). The footpoints of the reconnecting loops successively brightened and were observed as two flare ribbons in AIA 1600 {\AA} (see Figure 2(b)). The newly formed low-lying loops appeared as the post-flare loops in AIA 94 {\AA} (Figure 2(c)). Since the flux rope erupted, the filament material was ejected outward. This eruption process is consistent with the classical model of successful filament eruption associated with CMEs (e.g., Shibata et al. 1995; Lin \& Forbes 2000).

The filament in the first event was located in a sheared dipolar region embraced by the ambient magnetic fields with one polarity, which is also revealed in the $Q$ map. This kind of photospheric magnetic field often corresponds to the fan-spine configuration in the corona, which is favorable for the formation of circular flares (e.g., Masson et al. 2009; Wang \& Liu 2012; Sun et al. 2013; Zhang et al. 2016; Hao et al. 2017; Hernandez-Perez et al. 2017; Xu et al. 2017). The mini-filament in the second event had the similar morphology with the first filament. Most cases of this type of small-scale eruption were often observed in coronal holes, forming blowout jets associated with jet-like CMEs (Moore et al. 2010; Shen et al. 2012). It implies that this kind of filament eruptions may be ubiquitous on the Sun, from active regions with large flares to the quiet Sun with micro-flares.

According to the previous studies (e.g., Ji et al . 2003; Cheng et al. 2015), for a failed filament eruption, the rising filament will reach a peak height and then fall back to its initial location. However, the observed filaments in the present study erupted outward directly and moved along the large-scale lines continually, different from the behavior of failed filament eruptions. Therefore, the filaments erupted successfully instead of failed. Considering that the erupting filament material eventually reached the remote solar surface and no CME was produced, both the flares should be called confined flares. Our results reveal that some successful filament eruptions can trigger confined flares.

\acknowledgments { We are grateful to the anonymous referee for the valuable comments. We thank Dr. Qi Hao at Nanjing University and Dr. Yongliang Song at Peking University for helpful discussion. The data are used courtesy of ONSET and SDO science teams. This work is supported by the National Natural Science Foundations of China (11673035, 11790304, 11533008, 11790300), Key Programs of the Chinese Academy of Sciences (QYZDJ-SSW-SLH050), and the Youth Innovation Promotion Association of CAS (2014043).\\ }

{}

\clearpage

\end{document}